\documentclass{aa}
\usepackage{psfig,times}
\usepackage{amssymb}
\usepackage{graphicx}

\begin{document}
\title{Mean-field dynamo action in protoneutron stars}

\author{A.~Bonanno$^{1,2}$, L.~Rezzolla$^{3,4}$, and V.~Urpin$^{5,6}$}

\institute{$^{1}$ INAF, Osservatorio Astrofisico di Catania,
        Catania, Italy \\
        $^{2}$ INFN, Sezione di Catania, 
	Catania, Italy \\
	$^{3}$ SISSA, International School for Advanced Studies, 
	Trieste, Italy \\
        $^{4}$ INFN, Sezione di Trieste, 
        Trieste, Italy \\
	$^{5}$ A.F. Ioffe Institute of Physics and Technology, 
	St. Petersburg, Russia \\
	$^{6}$ Isaac Newton Institute of Chile in Eastern Europe and Eurasia, 
	St. Petersburg, Russia}

\titlerunning{Mean-field dynamo action in PNSs}
\authorrunning{Bonanno, Rezzolla and Urpin}
\date{\today}

\abstract{We have investigated the turbulent mean-field dynamo action in
	protoneutron stars that are subject to convective and neutron
	finger instabilities. While the first one develops mostly in the
	inner regions of the star, the second one is favoured in the
	outer regions, where the Rossby number is much smaller and a
	mean-field dynamo action is more efficient. By solving the
	mean-field induction equation we have computed the critical spin
	period below which no dynamo action is possible and found it to
	be $\sim 1$ s for a wide range of stellar models. Because this
	critical period is substantially longer than the characteristic
	spin period of very young pulsars, we expect that a mean-field
	dynamo will be effective for most protoneutron stars.

 }

\titlerunning{Mean-field dynamo action in PNSs}

\maketitle

\section{Introduction}
 
	The origin of the strong magnetic fields in neutron stars is
still a matter of controversy. The magnetic fields inferred from the
pulsar spin-down data ranges from $\sim 5 \times 10^{13}$ to $\sim
10^{8}$G but these values are representative of the global magnetic
configuration rather than of the fine magnetic structure near the stellar
surface. X-ray spectra of some pulsars have recently started to provide a
closer look at the magnetic field strength near the neutron star surface
and the absorption features in the spectrum of 1E 1207.4-5209, for
instance, have been used to estimate a strong surface magnetic field,
$B_{s} \sim 1.5 \times 10^{14}$G (Sanwal et al. 2002). This is to be
contrasted with the dipolar magnetic field estimated from the spin-down
rate of this pulsar: $B_{d} \sim (2-4) \times 10^{12}$G (Pavlov et
al. 2002), a value rather typical for a radio pulsar of $\sim 0.2-1.6$
Myr. As another example, Becker et al. (2002) have reported the presence
of an emission line in the X-ray spectrum of PSR B1821-24 that could be
interpreted as cyclotron emission from a corona above the pulsar's polar
cap. The line would be formed in a magnetic field $B_{s} \sim 3 \times
10^{11}$G, approximately two orders of magnitude stronger than the
dipolar magnetic field inferred from the spin evolution. Both
measurements provide evidence that the local magnetic fields at the
neutron star surface can be well above the dipolar one responsible for
the secular spin-down of pulsars.

	Observations of radio emitting pulsars also exhibit a distinction
between the dipole and surface magnetic fields. Recently, Gil \& Mitra
(2001) and Gil \& Melikidze (2002) have argued that the formation of a
vacuum gap in radio pulsars is possible if the actual surface magnetic
field near the polar cap is very strong, $B_{s} \sim 10^{13}$G,
irrespective of the magnetic field measured from the spin evolution.
Furthermore, the presence of a strong magnetic field with a small
curvature ($<10^{6}$ cm) can account for the radio emission of many
radiopulsars that lie in the pulsar graveyard and should be radio silent
(Gil \& Mitra 2001).
This growing number of evidences for a complex structure of the magnetic
field at the surface of neutron stars suggests that this may represent a
general property of pulsars.

	Magnetic fields with different strengths on different
length-scales can be explained naturally if they are generated through a
dynamo mechanism driven by turbulent motions. Indeed, it is generally
accepted that protoneutron stars (PNSs) are subject, shortly after their
birth, to hydrodynamic instabilities involving convective motions
(Epstein 1979, Livio et al. 1980, Burrows \& Lattimer 1986) and that
these can last $\sim 30-40$ s (Miralles et al. 2000, 2002). We here show
that, under suitable conditions, turbulent motions can generate magnetic
field via dynamo action and that a mean-field dynamo can be operative
together with small-scale dynamo processes (Thompson \& Duncan 1993; Xu
\& Busse 2001).


\section{Convection in PNSs}

	Hydrodynamic instabilities in PNSs can be driven by either lepton
gradients (Epstein 1979) leading to the so-called ``neutron-finger
instability'' (Bruenn \& Dineva 1996), or by negative entropy gradients
which commonly observed in simulations of supernova explosions (Bruenn \&
Mezzacappa 1994, 1995; Rampp \& Janka 2000) and evolutionary models of
PNSs (Keil \& Janka 1995; Keil et al. 1996; Pons et al. 1999). This
latter instability is usually referred to as the ``convective
instability'', although {\it both} instabilities involve convective
motions.

	It has been calculated that these instabilities will first
develop in the outer layers containing $\sim 30\%$ of the stellar mass,
with the convectively unstable region surrounded by the neutron-finger
unstable region, the latter involving therefore a larger portion of the
stellar material. After a few seconds the two unstable regions move
towards the inner parts of the star and after $\sim 10$ s from the
initial development, more than $90\%$ in mass of the star is
hydrodynamically unstable. At this stage the stellar core has become
convectively unstable but it is still surrounded by an extended
neutron-finger unstable region. In the $\sim 20 $ s that follow, the
temperature and lepton gradients are progressively reduced and the two
unstable regions begin to shrink, leaving the outer regions of the
star. After $\sim 30$ s, most of the PNS is stable and the instabilities
disappear completely after $\sim 40$ s (Miralles et al. 2000).

	During this period the PNS is opaque to neutrinos and the
turbulent mean velocity can be estimated within the mixing-length
approximation. The largest unstable length-scale is then of the order of
the pressure length-scale, $L \equiv p |dp/dr|^{-1}$, and the effective
flow velocity in this scale, $v_{_{L}}$, can be estimated as $v_{_{L}}
\approx L / \tau_{_{L}}$, where $\tau_{_{L}}$ is the growth-time of
instability which is of the order of the turnover time in the scale $L$
(Schwarzschild 1958). In some cases, $v_{_{L}}$ can be computed by
equating $\rho v_{_{L}}^{3}$, with $\rho$ the rest-mass density, to the
total neutrino energy flux (Thompson \& Duncan 1993). While the two
approaches yield similar results at the peak of the convective
instability, the first one is expected to be more accurate especially
when not all of the energy is transported by turbulence (e.g. when the
instability is not fully developed), when the temperature gradient is not
significantly super-adiabatic, or when the lepton gradient is small.

	In the {\it convectively unstable} region, the growth-time of
instability is
\begin{equation}
\frac{1}{\tau_{_{L}}^{2}} \sim \frac{1}{\tau_{c}^{2}} \sim \frac{1}{3}g 
\beta \frac{|\Delta \nabla T|}{T}\ ,
\end{equation}
where $\tau_{c}$ is the growth-time of convection, $g$ is the
gravitational acceleration, $\Delta \nabla T$ is the difference between
the actual and the adiabatic temperature gradient, and $\beta$ is the
coefficient of thermal expansion. Except during the last stages of the
unstable phase, when the entropy gradients have been washed out, the
convective instability grows on a short dynamical timescale. Miralles et
al. (2000), have estimated this to be $\tau_{c}\sim 0.1-1$ ms, from which
we derive the a mean turbulent velocity in the convectively unstable
region $v_{_{L}} \sim 10^{8}-10^{9}$ cm s$^{-1}$.

	In the {\it neutron-finger unstable} region, on the other hand,
the lepton number gradients dominate over the temperature gradients and
we can estimate the growth-time as
\begin{equation}
\frac{1}{\tau_{_{L}}^{2}} \sim \frac{1}{\tau_{nf}^{2}} \sim 
\frac{1}{3} g \delta | \nabla Y|\ ,
\end{equation}
where $\delta$ is the coefficient of chemical expansion, and $Y=(n_{e} +
n_{\nu})/n$ is the lepton fraction with $n_{e}$, $n_{\nu}$, and $n$ being
the number density of electrons, neutrinos, and baryons, respectively.
The estimated growth-time in the neutron-finger unstable region is a
couple of orders of magnitude {\it longer} than the one for the
convective instability, i.e. $\tau_{nf}\sim 30-100$ ms (Miralles et
al. 2000), thus yielding a mean turbulent velocity $v_{_{L}} \sim (1-3)
\times 10^{6}$ cm s$^{-1}$.

	The existence of two unstable regions with substantially
different mean velocities is the most significant difference with the
model by Thompson \& Duncan (1993) in which the whole PNS was assumed to
be convectively unstable with turbulent velocity $v_{_{L}} \sim 10^{8}$
cm s$^{-1}$. Furthermore, a longer growth-time (and hence turnover time)
is what promotes the efficiency of the mean-field dynamo in the
neutron-finger unstable region. We recall, in fact, that PNSs are likely
to rotate and although the initial spin rates of pulsars are not well
constrained by observations, they are believed to be around $\sim 100$ ms
(Narayan 1987). As a result, the Rossby number, $Ro = P/ \tau_{_{L}}$,
with $P$ being the PNS spin period, can take substantially different
values in the two unstable regions. In particular, $Ro \sim 100$ in the
convectively unstable region and the influence of rotation on the
turbulence is therefore weak. As a consequence, and as already pointed
out by Thompson \& Duncan (1993), the mean-field dynamo will not operate
efficiently here. In the more external regions unstable to
neutron-fingers, on the other hand, $Ro \sim 1$, turbulence can be
strongly modified by rotation and this favours the efficiency of a
mean-field dynamo. Of course, in both regions turbulent magnetic fields
can also be generated by small-scale dynamo driven by turbulent motions.

        To investigate more quantitatively the efficiency of a mean-field
dynamo action, we have modelled the PNS as a sphere of radius $R$ having
two spherical turbulent zones with substantially different properties and
separated at $R_{c}$. The inner parts ($r<R_{c}$) correspond to the
convectively unstable region, while the outer ones ($R_{c}<r<R$) to the
neutron-finger unstable region. The boundary between the two regions
moves inward on a timescale comparable to the cooling timescale
(i.e. $\sim 1-10$ s), much longer than the turnover time for both
instabilities. Hence $R_{c}$ is a slowly varying variable and it is
sufficient to consider a sequence of PNS models differing only for
$R_{c}$.

\section{Dynamo action in PNSs}

	Numerical simulations indicate that the turbulence in PNSs will
be non-stationary, developing rapidly soon after the collapse, reaching a
quasi-stationary regime after a few seconds, and then progressively
disappearing as the temperature and lepton gradients are removed. Because
the characteristic cooling timescale for the PNS exceeds both $\tau_{nf}$
and $\tau_{c}$, the turbulence can be treated adiabatically (this
assumption will cease to be accurate as the instabilities are
progressively suppressed). In this case, the mean-field induction
equation for a turbulent, magnetised and conducting plasma can be written
as
\begin{equation}
\label{ie}
\frac{\partial \vec{B}}{\partial t} = \nabla \times (\vec{v} 
	\times \vec{B} + \alpha \vec{B})
	- \nabla \times (\eta \nabla \times \vec{B}) \ ,
\end{equation}
where $\eta$ is the turbulent magnetic diffusivity, $\alpha$ is a
pseudo-scalar measuring the efficiency of the dynamo (the
``$\alpha$-parameter''). Here, $\vec{v}$ is the velocity the ordered
fluid motion, which we assume to follow a simple law, $\vec{v} =
\vec{\Omega} \times \vec{r}$, but allow for the differential rotation
often observed in numerical simulations (Zwerger \& M\"{u}ller 1997;
Rampp et al. 1998; Dimmelmeier et al. 2001). As customary in dynamo
theory, we express this differential rotation in terms of the radial
coordinate $r$ and through a simple quadratic law, i.e.
\begin{equation}
\label{drot}
\Omega(r) = \Omega_{_{0}} + r^{2} \Omega_{_{1}} \ ,
\end{equation}
where the coefficients $\Omega_{_{0}} \equiv \Omega(r=0)>0$ and
$\Omega_{_{1}}$ are not necessarily chosen so as to satisfy the Rayleigh 
stability criterion. Boundary conditions for the magnetic field need to be
specified at the stellar surface, where we impose vacuum boundary
conditions, and at the centre of the star, where we impose the vanishing
of the toroidal magnetic field.

	As discussed in Section 2, all of the PNS undergoes turbulent
motions but with properties that are different in the inner parts ($0
\leq r \lesssim R_{c}$), where fast convection operates, from those in
the outer parts ($R_{c} \lesssim r \leq R$), where the neutron-finger
instability operates. To model this in a simple way, we assume the
relevant physical properties of the two regions to vary in a smooth way
mostly across a thin layer of thickness $\Delta R = 0.025 R$. More
precisely, we express $\eta$ as
\begin{equation}
\label{eta}
\eta = \eta_c + (\eta_{nf}-\eta_c) \left\{1+
	erf[(r-R_c)/\Delta R] \right\}/2 \ ,
\end{equation}
where $\eta_{c}$ and $\eta_{nf}$ are respectively the turbulent magnetic
diffusivities caused by the convective and neutron-fingers instabilities,
and erf is the ``error function''. Using (\ref{eta}), $\eta \approx
\eta_{c}$ in the convectively unstable zone ($R_{c} - r \gg \Delta R$),
while $\eta \approx \eta_{nf}$ in the neutron-finger unstable zone
($r-R_{c} \gg \Delta R$). Furthermore, because in typical PNSs $\eta_{nf}
\ll \eta_{c}$, we have here chosen $\eta_{nf}/\eta_c = 0.1$. Similarly,
we have modelled the $\alpha$-parameter as being negligibly small in the
convectively unstable region and equal to $\alpha_{nf}$ in the
neutron-finger unstable region, i.e.
\begin{equation}
\label{alp}
\alpha(r,\theta) = \alpha_{nf}\cos\theta \left\{1 + 
	erf[(r-R_c)/{\Delta R}] \right\} /2 \ , 
\end{equation}
where the angular dependence chosen in (\ref{alp}) is the simplest
guaranteeing antisymmetry across the equator.

	We recall that in a rotating turbulence with lengthscale $\ell$
and moderate Rossby number, $\alpha \approx - \Omega \ell^{2} \nabla \ln
(\rho v_{_{L}}^{2})$ (R\"{u}diger \& Kitchatinov 1993). In PNSs, however,
the pressure is determined by degenerate neutrons and can, in a first
approximation, be expressed with a simple relation of the type $p \propto
\rho^{\gamma}$, where $\gamma=5/3$ for a non-relativistic neutron gas and
$\gamma=4/3$ for a relativistic one. As a result, the density lengthscale
is comparable to the pressure one $L$ and, as mentioned in Section 2, to
the lengthscale of the instabilities. This introduces a great
simplification since we can express the isotropic turbulence in the
neutron-finger unstable zone simply as $\alpha_{nf} \approx \Omega L$.

\section{Numerical results}

	The induction equation (\ref{ie}) with $\eta$ and $\alpha$ given
by (\ref{eta}) and (\ref{alp}) has been solved with a numerical code
employing finite-difference techniques for the radial dependence and a
polynomial expansion for the angular dependence (see Bonanno et al. 2002
for details). The simulations reported here use 30 spherical harmonics
and about 40 grid points in the radial direction; details on the
numerical procedure will appear elsewhere.

	Assuming $v_{_{L}} L/3 = 10^{11}$ cm$^2$ s$^{-1}$ and
progressively varying the velocity field (\ref{drot}) in terms of the
differential rotation parameter $q \equiv R^{2}
\Omega_{_{1}}/\Omega(r=R)$, we have solved equation (\ref{ie}) to
determine the critical value $\alpha_{_{0}}$ corresponding to the
marginal stability of the dynamo. Given a certain amount of differential
rotation, in fact, the seed magnetic field will grow if $\alpha_{nf} >
\alpha_{_{0}}$ and instead decay if $\alpha_{nf} < \alpha_{_{0}}$.  The
different types of dynamo can be distinguished according to whether the
differential rotation is small and the evolution of the magnetic field
stationary (i.e. a $\alpha^2$-dynamo), or viceversa (i.e. a
$\alpha\Omega$-dynamo).

\begin{figure}
\begin{center}
\includegraphics[width=9.0cm]{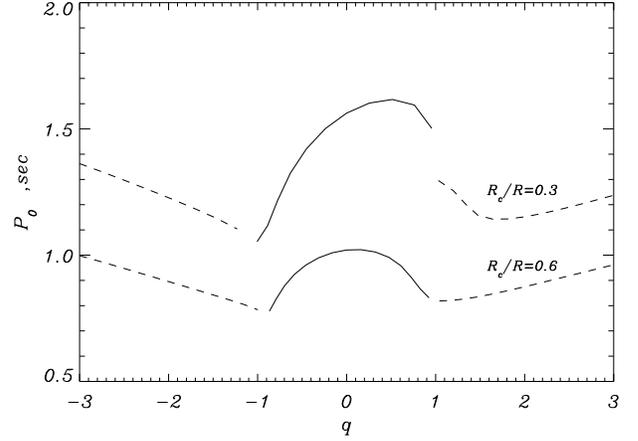}
\caption{Critical period as a function of the differential rotation
parameter. The two curves refer to different values of $R_{c}$, with the
solid parts corresponding to a $\alpha^{2}$-dynamo and the dashed parts
to a $\alpha \Omega$-dynamo.}
\end{center}
\end{figure}

	Since $\alpha_{nf} \approx \Omega L$, the critical value
$\alpha_{0}$ effectively selects a critical value for the spin period,
$P_{_{0}} \equiv 2 \pi L/ \alpha_{0}$, such that magnetic field
generation via a mean-field dynamo action will possible only if the
stellar spin period is shorter than the critical one. In Fig.~1, we plot
the critical period as a function of the differential rotation parameter
$q$. Note that $q < 0$ and $q > 0$ correspond to situations in which the
stellar surface rotates faster and slower than the centre, respectively
(values $q < -1$ correspond to a counter-rotation and may be not
physically relevant).

	As shown in Fig.~1, a stationary $\alpha^{2}$-dynamo dominates
the magnetic field generation process for $|q| < 1$, while a $\alpha
\Omega$-dynamo is more efficient for $|q| > 1$. This latter case will be
characterised by a magnetic field of oscillating strength but, given the
large differential rotation required, it may be difficult to achieve in
practice. Hence, the $\alpha^{2}$-dynamo appears to be the most likely
source of magnetic field generation via dynamo processes in PNSs.

	The critical spin found here is in general rather long and for a
PNS rotating uniformly (i.e. $q=0$), a mean-field dynamo will develop if
$P \leq 1$ s when $R_{c}/R =0.6$ and if $P \leq 1.5$ s when $R_{c}/R
=0.3$. This difference is due to the fact that a PNS with a more extended
neutron-finger unstable region can rotate proportionally more slowly
while maintaining the same dynamo action. Furthermore, if the star
rotates differentially with $|q| \approx 1$, $P_{_{0}}$ is further
reduced, being approximately 20\% shorter. As a result, only PNSs with $P
\gtrsim 1-1.5$ s {\it will not develop} a turbulent mean-field dynamo.
Such slow rotation rates should be rather difficult to achieve if angular
momentum is conserved during the collapse to a PNS and, indeed,
observations suggest that the initial periods of pulsars are considerably
shorter than the critical period $P_{_{0}}$ obtained here. We expect,
therefore, that a turbulent mean-field dynamo will be effective during
the initial stages of the life of most PNSs.

\begin{figure}
\includegraphics[width=9.0cm]{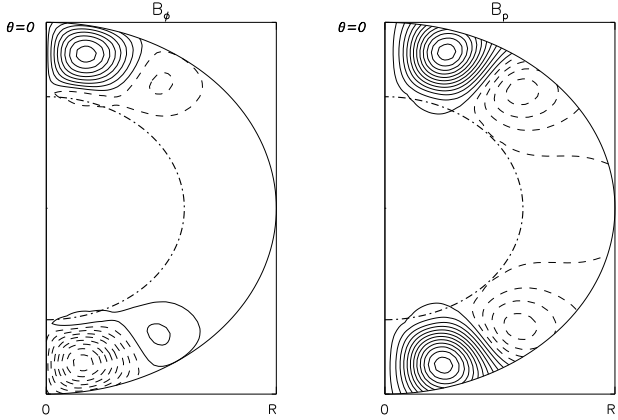}
\caption{Toroidal ($B_{\phi}$) and the poloidal ($B_p$) magnetic field
lines. Solid and dashed contours correspond to positive and negative
values, respectively. The dot-dashed line marks the position of $R_c =
0.6 R$.}
\end{figure}

	In Fig.~2 we show the toroidal ($B_{\phi}$) and poloidal
($B_{p}$) magnetic fields of a typical PNS model. Note that both
components are generated in the outer neutron-finger unstable region,
but turbulent diffusion produces a magnetic field also in the inner
convectively unstable region, although this is considerably weaker. The
toroidal magnetic field tends to concentrate near the polar regions
whereas the poloidal one is more evenly distributed in latitude. The
numerical calculations also show that if $|q|<1$ and the field is
generated by the $\alpha^{2}$-dynamo then $B_{\phi}/B_{p} \sim 10$, while
$B_{\phi}/B_{p} \sim 100-200$ if $|q|>1$ and the $\alpha \Omega$-dynamo
generates the magnetic field. Both results suggest that the internal
magnetic fields in neutron stars could be substantially stronger than the
observable surface ones.

\section{Conclusions}

	We have calculated the turbulent mean-field dynamo during the
turbulent instabilities that are expected to accompany the early stages
of the life of a PNS. For $\sim 30-40$ s, in fact, the PNS is subject to
two substantially different instabilities, with a convective instability
active in the inner regions of the star and a neutron-finger instability
being more efficient in the outer regions. The turbulent motions are more
rapid in the convectively unstable zone, where the Rossby number is large
and the $\alpha$-parameter is likely not to be significant. In the
neutron-finger unstable region, on the other hand, the turnover time is
considerably longer, the Rossby number small, and the $\alpha$-parameter
is sufficiently large that a mean-field dynamo can operate.

	The occurrence of a dynamo depends sensitively on the stellar
rotation rate since this can influence the turbulent motions if
sufficiently high. Our simulations show that even relatively slowly
rotating PNSs can be subject to a dynamo action in the neutron-finger
unstable region, with the $\alpha^2$-dynamo being the most efficient
mechanism. The critical value of the spin period below which the dynamo
is suppressed has been found to be $P_{_{0}} \sim 1 $ s for a wide range
of models and is essentially larger than the characteristic spin period
of young pulsars as inferred from observations. As a result, a turbulent
mean-field dynamo can be effective in the early stages of the life of
most PNSs. The magnetic field produced in this way is concentrated mostly
where the generation occurs, but turbulent diffusivity transports part of
it also in the inner regions of the star.

	We have here considered the generation of axisymmetric magnetic
fields, but the $\alpha^{2}$-dynamo could also lead to the generation of
non-axisymmetric magnetic fields, with the critical spin period being not
very different from the one discussed in the axisymmetric case (R\"udiger
et al. 2003). The generation of non-axisymmetric magnetic fields and the
effect of saturation will be considered in a forthcoming paper.

	We note that the production of the mean magnetic field by a
turbulent dynamo is accompanied by the generation of small-scale magnetic
fields which are likely to be stronger than the mean one and that have
not been considered here. However, nonlinear effects associated to these
small-scale fields can influence the dynamo and need to be taken properly
into account.

\acknowledgements 
We thank the referee, A. Brandenburg, for his useful comments. Financial
support has been provided by the MIUR and the EU Network Program
(HPRN-CT-2000-00137). LR acknowledges hospitality at the KITP in Santa
Barbara (NSF grant PHY99-07949). VU thanks INFN (Catania) for hospitality
and support.

{}

\end{document}